\documentclass{ws-procs961x669}

\begin{document}

\title{Retarded potentials and radiation in odd dimensions}

\author{D. V. Gal'tsov$^a$ and M. Khlopunov$^{a,b}$}

\address{$^a$Faculty of Physics, Lomonosov Moscow State University,\\
Moscow, 119899, Russia\\
$^b$Institute of Theoretical and Mathematical Physics, Lomonosov Moscow State University,\\
Moscow 119991, Russia\\
E-mail: galtsov@phys.msu.ru, khlopunov.mi14@physics.msu.ru}

\begin{abstract}
Free massless fields of any spin in flat D-dimensional spacetime propagate at the speed of light. But the retarded fields produced by the corresponding point-like moving sources share this property only for even D. Since the Green’s functions of the d’Alembert equation are localized on the light cone in even-dimensional spacetime, but not in odd dimensions, extraction of the emitted part of the retarded field in odd D requires some care. We consider the wave equations for spins 0, 1, and 2 in five-dimensional spacetime and analyze the fall-off conditions for the retarded fields at large distances. It is shown that the farthest part of the field contains a component propagating at the speed of light, while the non-derivative terms propagate with all velocities up to that of light. The generated radiation will contain a radiation tail corresponding to the complete prehistory of the source’s motion preceding the retarded moment of time. We also demonstrate that dividing the Green’s function into a part localized on the light cone and another part that is not zero inside the light cone gives separately the divergent terms in the Coulomb field of a point source. Their sum, however, is finite and corresponds to the usual power-law behaviour.
\end{abstract}


\bodymatter

\section{Introduction}

The recent interest in the theory of radiation in spacetime dimensions other than four is mostly related to the development of the theories with extra dimensions of spacetime. While the superstring theory, pretending to the status of fundamental theory, predicts the existence of extra dimensions, there is a number of phenomenological multi-dimensional gravity theories \cite{Arkani-Hamed:1998jmv,Randall:1999ee,Randall:1999vf,Dvali:2000hr} solving some problems of elementary particles physics and cosmology. However, the characteristics of extra dimensions, such as their number, geometry and size, vary widely from one theory to another.

The actively developing gravitational-wave astronomy is the one of the most promising tools to probe the extra dimensions. So, the first constraints on the characteristics of extra dimensions have already been obtained by use of the GW170817/GRB170817A event data \cite{Visinelli:2017bny,Pardo:2018ipy,Chakravarti:2019aup}. The possibility of using the future gravitational-wave observatories, such as LISA, to constrain the extra dimensions on cosmological scales has also been discussed \cite{Deffayet:2007kf,Corman:2020pyr}. Also, it is worth to note the recent advances in the constraining extra dimensions by the photograph of the supermassive black hole M87* shadow \cite{Vagnozzi:2019apd,Banerjee:2019nnj}.

However, in most of the literature, only the radiation in even-dimensional spacetimes is considered \cite{Kosyakov:1999np,Cardoso:2002pa,Cardoso:2007uy,Mironov:2007nk,Kosyakov:2008wa}, while the odd dimensions have been mainly discussed in the context of radiation reaction force \cite{Galtsov:2001iv,Kazinski:2002mp,Yaremko:2007zz,Shuryak:2011tt}. It is mostly due to the Huygens principle violation in odd dimensions known since the classical works of Hadamard, Courant and Hilbert, Ivanenko and Sokolov \cite{Hadamard:book,Courant:book,Ivanenko:book}. In any dimensions, the signal from the instant flash of the source reaches an observer in the interval of time required for it to propagate at the speed of light. However, in odd dimensions, the endless tail signal decaying with time is observed after that, which is not the case in even dimensions. Mathematically, the Huygens principle violation consists in odd-dimensional retarded Green's functions being localised not only on the light cone, as they are in even dimensions, but also inside it. As a result, the retarded fields in odd dimensions propagate in space with all velocities up to that of light. However, free massless fields propagate exactly at the speed of light in any dimensions. Therefore, there is the apparent mismatch as the radiation being the free field far from the source is determined by its retarded field.

In this paper, we demonstrate that despite the Huygens principle violation in odd dimensions the radiation can be computed by the integration of the energy-momentum flux in the wave zone. We use the Rohrlich-Teitelboim radiation definition \cite{Rohrlich:1961,Rohrlich:2007,Teitelboim:1970} (see, also, \cite{Kosyakov:1992qx,Galtsov:2004uqu,Spirin:2009zz}) based on the Lorentz-invariant decomposition of the on-shell energy-momentum tensor. Considering the radiation of spin-0, spin-1 and spin-2 fields in five spacetime dimensions we show that the emitted part of the field energy-momentum propagates in space exactly at the speed of light, while it depends on the entire history of the source motion preceding the retarded time, in contrast with the four-dimensional theory.

The paper is organised as follows. In second section we consider the scalar radiation from the point charge in five spacetime dimensions. We briefly recall the recurrent relation for the odd-dimensional retarded Green's functions and the Rohrlich-Teitelboim approach to radiation. Based on the latter, we compute the emitted part of the on-shell energy-momentum tensor. In Sec. 3, we consider the analogous problem for the five-dimensional electromagnetic field. Section 4 is devoted to the calculation of the gravitational radiation from the point particle moving along an arbitrary world line in five dimensions. In the last section we briefly discuss our results.

\section{Scalar radiation in five dimensions}

Action of the massless scalar field $\varphi(x)$ interacting with the massive point particle moving along an arbitrary world line $z^{\mu}(\tau)$ in the five-dimensional Minkowski spacetime is written as
\begin{equation}
\label{eq:sc_action}
S = \frac{1}{4\pi^2} \int d^{4+1}x \, \partial^{\mu} \varphi(x) \partial_{\mu} \varphi(x) - \int d\tau \sqrt{ \dot{z}^{\alpha} \dot{z}_{\alpha} } ( m + g \varphi(z) ), \quad \dot{z}^{\mu} = \frac{dz^{\mu}}{d\tau},
\end{equation}
where $m$ is the particle's mass, $g$ its scalar charge, and Minkowski metric is $\eta_{\mu\nu} = {\rm diag}(1,-1,-1,-1,-1)$. We assume that particle's motion is governed by the external forces and is not affected by the radiation of scalar field, so here the world line variables are non-dynamical.

The action \eqref{eq:sc_action} leads to the scalar field equation of motion in form
\begin{align}
\label{eq:sc_EoM}
&\square \varphi(x) = - 2\pi^2 j(x), \\
\label{eq:sc_curr}
&j(x) = g \int d \tau \sqrt{ \dot{z}^{\alpha} \dot{z}_{\alpha} } \delta^{(4+1)} (x-z),
\end{align}
where $j(x)$ is the scalar current, and $\square = \partial^{\mu} \partial_{\mu}$ is the d'Alembert operator on the five-dimensional Minkowski space. To determine the radiation energy-momentum flux carried by the scalar field, we use its canonical energy-momentum tensor
\begin{equation}
\label{eq:sc_EMT}
T_{\mu\nu} (x) = \frac{1}{2\pi^2} \left( \partial_{\mu} \varphi \partial_{\nu} \varphi - \frac{1}{2} \eta_{\mu\nu} \partial^{\alpha} \varphi \partial_{\alpha} \varphi \right).
\end{equation}

The retarded solution of the Eq. \eqref{eq:sc_EoM} is given as
\begin{equation}
\label{eq:5D_sc_ret_gen}
\varphi(x) = - 2\pi^2 \int d^{4+1} x' \, j(x') G_{\rm ret}^{4+1} (x-x'), \\
\end{equation}
where $G_{\rm ret}^{4+1} (x)$ is the retarded Green's function of the five-dimensional d'Alembert equation. It is defined by the equation
\begin{align}
&\square G_{\rm ret}^{4+1} (x) = \delta^{(4+1)} (x),\\
&G_{\rm ret}^{4+1}(x)=0, \, x^0 < 0.
\end{align}
In the odd-dimensional Minkowski spacetimes, retarded Green's functions are determined by the following recurrent relation \cite{Ivanenko:book} (see, also, \cite{Galtsov:2020hhn})
\begin{equation}
\label{eq:odd_Green_recurr}
G^{2\nu+1}_{\rm ret}(x)=\frac{(-1)^{\nu-1}}{(2\pi)^{\nu}}\frac{d^{\nu-1}}{(rdr)^{\nu-1}}\frac{\theta(t)\,\theta(t^2-r^2)}{\sqrt{t^2-r^2}}, \; \nu \in \mathbb{N},
\end{equation}
where $t=x^0$ and $r=|\mathbf{x}|$. Considering the expression under derivatives in Eq. \eqref{eq:odd_Green_recurr} as a product of separate distributions and taking into that $d\theta(x)/dx=\delta(x)$, we find the five-dimensional retarded Green's function as
\begin{equation}
\label{eq:5D_Green}
G_{\rm ret}^{4+1} (x) = \frac{\theta (t)}{2 \pi^2} \left \lbrack \frac{\delta (t^2-r^2)}{(t^2-r^2)^{1/2}} - \frac{1}{2} \frac{\theta (t^2-r^2)}{(t^2-r^2)^{3/2}} \right \rbrack.
\end{equation}
As discussed above, it is localised not only on the light cone, but also inside it, leading to the propagation of the retarded field in space with all velocities up to that of light. Also, retarded field depends on the entire history of the source's motion preceding the retarded time and is given by the sum of separately divergent on the light cone $t^2-r^2=0$ terms. However, one can show that the divergences contained in each of the terms in Eq. \eqref{eq:5D_Green} mutually cancel out and the resulting field is finite.

\subsection{Field of a static charge}

Let us consider the case of static particle, to demonstrate the absence of divergences in the retarded field. In our calculations, we will use the "finite lifetime" trick to show the cancellation of divergences between the two terms.

We assume that the particle is at the origin and it is "switched on" for a finite interval of the proper time $a<\tau<b$, where $a<0$ and $b>0$ for concreteness. For this interval its world line has the form $z^{\mu} = \lbrack \tau, 0, 0, 0, 0 \rbrack$. Then, by use of Eqs. \eqref{eq:5D_sc_ret_gen}, \eqref{eq:sc_curr} and \eqref{eq:5D_Green} the scalar field is written as
\begin{equation}
\varphi  (x) = \frac{g}{2} \int_{a}^{b} d \tau \left \lbrack \frac{\theta (t-\tau-r-\epsilon)}{\lbrack (t-\tau)^2 - r^2 \rbrack^{3/2}} - \frac{\delta (t-\tau-r-\epsilon)}{r \lbrack (t-\tau)^2 - r^2 \rbrack^{1/2}} \right \rbrack,
\end{equation}
where we introduced the regularising parameter $\epsilon \to +0$ into the delta and Heaviside functions to shift the divergences from the light cone. Preforming the integration we obtain
\begin{equation}
\varphi  (x) = \frac{g}{2} \begin{cases} 0, \; t < a+r, \\
\displaystyle -  \frac{t-a}{r^2 \lbrack (t-a)^2 - r^2 \rbrack^{1/2}}, \; a+r \leq t < b+r, \\
\displaystyle  \frac{t-b}{r^2 \lbrack (t-b)^2 - r^2 \rbrack^{1/2}} - \frac{t-a}{r^2 \lbrack (t-a)^2 - r^2 \rbrack^{1/2}} , \; t \geq b+r.
\end{cases}
\end{equation}
In the limit of eternal particle, we arrive at the finite Coulomb-like field
\begin{equation}
\lim_{a,b \to \pm \infty} \varphi(x) = - \frac{g}{2r^2},
\end{equation}
with the power-law behaviour corresponding to the increased dimensionality of space. Similar cancellation of divergences has been shown to take place, also, for the moving particle \cite{Galtsov:2020hhn}.

\subsection{The Rohrlich-Teitelboim radiation definition}

The structure of the odd-dimensional Green's functions \eqref{eq:odd_Green_recurr} makes the extraction of the emitted part of the retarded field in a standard manner non-trivial and requires a more sophisticated approach.

Such an approach was suggested by Rohrlich \cite{Rohrlich:1961,Rohrlich:2007} and Teitelboim \cite{Teitelboim:1970} (see, also, \cite{Kosyakov:1992qx,Galtsov:2004uqu,Spirin:2009zz}). It is based on the use of certain covariantly defined quantities, so we briefly recall their definitions. Let us consider the point particle moving along the world line $z^{\mu}(\tau)$ with velocity $v^{\mu}=dz^{\mu}/d\tau$ in the $D$-dimensional spacetime. The observation point coordinates are denoted as $x^{\mu}$. Assume the observation point to be a top of the light cone in the past and denote the intersection point of this light cone with the particle's world line as $z^{\mu}(\hat{\tau}) \equiv \hat{z}^{\mu}$. The corresponding moment of proper time $\hat{\tau}$ is called the retarded proper time and is defined by equation
\begin{equation}
\label{eq:ret_prop_time_def}
(x^{\mu} - \hat{z}^{\mu})^2 = 0, \; x^0 \geq \hat{z}^0.
\end{equation}
Further, all the hatted quantities correspond to that moment. Then, we introduce three spacetime vectors: a lightlike vector $\hat{X}^{\mu} = x^{\mu} - \hat{z}^{\mu}$ directed from the retarded point of the world line to the observation point, a spacelike vector $\hat{u}^{\mu}$ orthogonal to the particle's velocity at the retarded moment of proper time, and a lightlike vector $\hat{c}^{\mu} = \hat{u}^{\mu} + \hat{v}^{\mu}$ aligned with the vector $\hat{X}^{\mu}$. Introduced vectors have the following properties
\begin{equation}
\hat{X}^2 = 0, \quad (\hat{u}\hat{v}) = 0, \quad \hat{u}^2 = - \hat{v}^2 = -1, \quad \hat{c}^2 = 0,
\end{equation}
where $(\hat{u}\hat{v}) \equiv \hat{u}^{\alpha}\hat{v}_{\alpha}$. Using these vectors we, also, introduce the Lorentz-invariant distance $\hat{\rho}$ being the scalar product of two spacetime vectors
\begin{equation}
\hat{\rho} \equiv (\hat{v}\hat{X}), \quad \hat{X}^{\mu} = \hat{\rho} \hat{c}^{\mu}.
\end{equation}
It is equal to the spatial distance in the Lorentz frame comoving with the particle at the retarded proper time. If the particle moves inside the compact region of space, then the Lorentz-invariant distance $\hat{\rho}$ is equivalent to the spatial distance
\begin{equation}
\hat{\rho} \xrightarrow{r \gg |\hat{\mathbf{z}}|} r.
\end{equation}
In accordance with the Rohrlich-Teitelboim approach, it is the Lorentz-invariant distance $\hat{\rho}$ that is used in the long-range expansion of tensors and definition of the wave zone.

In the Rohrlich-Teitelboim approach, the radiation is determined by the most long-range part of the on-shell energy-momentum tensor expansion in the inverse powers of the Lorentz-invariant distance $\hat{\rho}$. In $D$ dimensions, the retarded field's on-shell energy-momentum tensor is expanded as \cite{Teitelboim:1970,Kosyakov:1992qx,Kosyakov:1999np,Galtsov:2004uqu,Kosyakov:2008wa,Spirin:2009zz}
\begin{align}
&T^{\mu\nu} = T^{\mu\nu}_{\rm Coul} + T^{\mu\nu}_{\rm mix} + T^{\mu\nu}_{\rm rad} \\
&T^{\mu\nu}_{\rm Coul} \sim \frac{A^{\mu\nu}}{\hat{\rho}^{2D-4}}, \quad T^{\mu\nu}_{\rm mix} \sim \frac{B^{\mu\nu}}{\hat{\rho}^{2D-5}} + \ldots + \frac{C^{\mu\nu}}{\hat{\rho}^{D-1}}, \quad T^{\mu\nu}_{\rm rad} \sim \frac{D^{\mu\nu}}{\hat{\rho}^{D-2}}.
\end{align}
Here, the first term $T^{\mu\nu}_{\rm Coul}$ is the energy-momentum tensor of the deformed Coulomb-like part of the retarded field. The second one is the mixed part, which consists of more than one term for $D>4$ and is absent in $D=3$. The most long-range part $T^{\mu\nu}_{\rm rad}$ of the on-shell energy-momentum tensor expansion has the properties allowing to associate it with the radiation energy-momentum:
\begin{itemize}
\item
It is separately conserved $\partial_{\mu} T^{\mu\nu}_{\rm rad} = 0$, corresponding to its dynamical independence from the other parts;
\item
It is proportional to the direct product of two null vectors $T^{\mu\nu}_{\rm rad} \sim \hat{c}^{\mu}\hat{c}^{\nu}$, corresponding to its propagation exactly with the speed of light $\hat{c}_{\mu} T_{\rm rad}^{\mu\nu} = 0$;
\item
It falls down as $T^{\mu\nu}_{\rm rad} \sim 1/r^{D-2}$ and gives positive definite energy-momentum flux through the distant $(D-2)$-dimensional sphere of area $\sim r^{D-2}$.
\end{itemize}
Therefore, the radiation power in $D$-dimensions can be computed as the energy flux associated with $T^{\mu\nu}_{\rm rad}$ through the distant $(D-2)$-dimensional sphere of radius $r$
\begin{equation}
W_{D} = \int \, T_{\rm rad}^{0 i} \; n^{i}\, r^{D-2} \, d\Omega_{D-2}; \quad i = \overline{1, D-1},
\end{equation}
where $n^i = x^i/r$ is the unit spacelike vector in the direction of observation, and $d\Omega_{D-2}$ is the angular element on the $(D-2)$-dimensional sphere. This structure holds in both even and odd dimensions with the only difference that in odd dimensions the emitted part of the energy-momentum tensor depends on the entire history of the particle's motion preceding the retarded proper time $\hat{\tau}$.

Note that, usually, the energy-momentum tensor is just the bilinear form of the field derivatives, as, e.g., in Eq. \eqref{eq:sc_EMT}. Then, one can define the emitted part of the retarded field derivative, by analogy with that of the energy-momentum tensor,
\begin{equation}
(\partial_{\mu} \Phi)^{\rm rad} \sim 1/\hat{\rho}^{D/2-1}.
\end{equation}
This definition is valid in the cases of scalar and electromagnetic fields and linearised gravity considered below.

\subsection{Emitted part of the scalar field}

Now we turn to the calculation of the emitted part of the retarded scalar field derivative. The computations below are similar to that in \cite{Galtsov:2020hhn}, where we have considered the scalar synchrotron radiation in three and five spacetime dimensions.

Using the Eq. \eqref{eq:5D_sc_ret_gen} together with Eqs. \eqref{eq:5D_Green} and \eqref{eq:sc_curr} we obtain the retarded scalar field of the moving particle as
\begin{equation}
\varphi(x) = - g \int d\tau \, \theta(X^0(\tau)) \left \lbrack \frac{\delta(X^2(\tau))}{(X^2(\tau))^{1/2}} - \frac{1}{2} \frac{\theta(X^2(\tau))}{(X^2(\tau))^{3/2}} \right \rbrack,
\end{equation}
where we introduced the vector $X^{\mu}(\tau) = x^{\mu} - z^{\mu}(\tau)$. In what follows, we omit its dependence on the proper time for brevity.

Its derivative is found to have the form
\begin{equation}
\partial_{\mu} \varphi(x) = - 2g \int d\tau \, \theta(X^{0}) \left \lbrack \frac{3}{4} \frac{\theta(X^{2})}{(X^{2})^{5/2}} + \frac{\delta^{\prime}(X^{2})}{(X^{2})^{1/2}} - \frac{\delta(X^{2})}{(X^{2})^{3/2}} \right \rbrack X_{\mu},
\end{equation}
where $\delta'(x) = d\delta(x)/dx$. Integrating by parts the term containing derivative of delta function by use of the relation
\begin{equation}
\label{eq:delta_int_parts}
\frac{dX^{2}}{d\tau} = - 2(vX),
\end{equation}
we obtain the scalar field derivative in form
\begin{multline}
\label{eq:5D_scal_deriv_gen}
\partial_{\mu} \varphi(x) = - g \int d\tau \, \theta(X^{0}) \left \lbrack \frac{3}{2} \frac{\theta(X^{2})}{(X^{2})^{5/2}} X_{\mu} - \frac{\delta(X^{2})}{(X^{2})^{3/2}} X_{\mu} - \right. \\ \left. - \frac{\delta(X^{2})}{(vX)^{2} (X^{2})^{1/2}} \left \lbrack (aX) - 1 \right \rbrack X_{\mu} - \frac{\delta(X^{2})}{(vX) (X^{2})^{1/2}} v_{\mu} \right \rbrack,
\end{multline}
where we introduced the acceleration vector $a^{\mu}=d^2 z^{\mu}/d\tau^2$.

To obtain its emitted part, we extract the leading $\hat{\rho}$-asymptotic of the Eq. \eqref{eq:5D_scal_deriv_gen}. We transform the products of delta and Heaviside functions by use of the relation for the delta function of complex argument
\begin{equation}
\label{eq:delta_compl_arg}
\theta(X^{0}) \delta(X^{2}) = \frac{\delta(\tau - \hat{\tau})}{2\hat{\rho}}.
\end{equation}
We, also, rewrite the vector $X^{\mu}$ in terms of the Lorentz-invariant distance
\begin{equation}
\label{eq:X_rho_expan}
X^{\mu} = Z^{\mu} + \hat{\rho} \hat{c}^{\mu}, \quad Z^{\mu} = \hat{z}^{\mu} - z^{\mu}.
\end{equation}
Then, by use of Eqs. \eqref{eq:delta_compl_arg} and \eqref{eq:X_rho_expan} we arrive at the emitted part of the scalar field derivative in form
\begin{equation}
\label{eq:5D_sc_der_emit_unreg}
( \partial_{\mu} \varphi(x) )^{\rm rad} = - \frac{g \hat{c}_{\mu}}{2^{5/2} \hat{\rho}^{3/2}} \int_{-\infty}^{\hat{\tau}} d\tau \left \lbrack \frac{3}{2} \frac{1}{(Z\hat{c})^{5/2}} - \frac{\delta(\tau - \hat{\tau})}{(Z\hat{c})^{3/2}} - \frac{2 (a\hat{c})\delta(\tau - \hat{\tau})}{(v\hat{c})^{2} (Z\hat{c})^{1/2}} \right \rbrack,
\end{equation}
where we have taken into account that up to the leading order $X^{2} \sim 2 \hat{\rho} (Z\hat{c})$ and $(vX) \sim \hat{\rho} (v\hat{c})$. Note that each integral in the Eq. \eqref{eq:5D_sc_der_emit_unreg} diverges on the upper integration limit $\tau \to \hat{\tau}$, as $(Z\hat{c}) \to (\hat{v}\hat{c}) (\hat{\tau} - \tau) = (\hat{\tau} - \tau)$. However, one can show that the last two terms of the integrand do not carry physical information concerning the field in the wave zone and just eliminate the divergences contained in the first one. To do this, we introduce the regularising parameter $\epsilon \to +0$ into the argument of delta function and perform integration by use of it, obtaining the divergent result
\begin{equation}
\label{eq:5D_sc_c-t_1}
\int_{-\infty}^{\hat{\tau}} d\tau \, \frac{\delta(\tau - \hat{\tau} + \epsilon)}{(Z\hat{c})^{3/2}} = \frac{1}{\epsilon^{3/2}},
\end{equation}
which can be rewritten as
\begin{equation}
\label{eq:5D_sc_c-t_2}
\frac{1}{\epsilon^{3/2}} = \frac{3}{2} \int_{-\infty}^{\hat{\tau}-\epsilon} \frac{d\tau}{(\hat{\tau} - \tau)^{3/2}}.
\end{equation}
By analogy, the second term with the delta function transforms to
\begin{equation}
\label{eq:5D_sc_c-t_3}
\int_{-\infty}^{\hat{\tau}} d\tau \, \frac{ (a\hat{c}) \delta(\tau - \hat{\tau})}{(v\hat{c})^{2} (Z\hat{c})^{1/2}} = \frac{1}{2} \int_{-\infty}^{\hat{\tau}-\epsilon} d\tau \frac{(\hat{a}\hat{c})}{(\hat{\tau} - \tau)^{3/2}}.
\end{equation}
Then, omitting the regularising parameter in the upper integration limit we find the emitted part of the scalar field derivative as
\begin{equation}
\label{eq:5D_sc_emit}
( \partial_{\mu} \varphi(x) )^{\rm rad} = - \frac{g \hat{c}_{\mu}}{2^{5/2} \hat{\rho}^{3/2}} \int_{-\infty}^{\hat{\tau}} d\tau \left \lbrack \frac{3}{2} \frac{1}{(Z\hat{c})^{5/2}} - \frac{3}{2} \frac{1}{(\hat{\tau} - \tau)^{5/2}} - \frac{(\hat{a}\hat{c})}{(\hat{\tau}-\tau)^{3/2}} \right \rbrack.
\end{equation}
In practice, calculating the radiation from the particle moving along some trajectory one needs to make the convergence of the integral in Eq. \eqref{eq:5D_sc_emit} explicit. To achieve it, one has to reintroduce the regularising parameter $\epsilon \to +0$ into the upper integration limit and perform some transformation of the first term in the integrand: usually, it is the integrations by parts, which extract the divergences from it and cancel them out by use of the remaining two integrals. The above structure of the emitted part of the field holds in any odd spacetime dimensions \cite{Galtsov:2020hhn}.

Substituting the emitted part of the scalar field derivative into the Eq. \eqref{eq:sc_EMT} we find the radiated part of the energy-momentum tensor
\begin{align}
\label{eq:sc_EMT_emit_1}
&T_{\mu\nu}^{\rm rad} (x) = \frac{g^2 \hat{c}_{\mu} \hat{c}_{\nu}}{64 \pi^2 \hat{\rho}^3} {\cal A}^{2} (x), \\
\label{eq:sc_EMT_emit_2}
&{\cal A}(x) = \int_{-\infty}^{\hat{\tau}} d\tau \left \lbrack \frac{3}{2} \frac{1}{(Z\hat{c})^{5/2}} - \frac{3}{2} \frac{1}{(\hat{\tau}-\tau)^{5/2}} - \frac{(\hat{a}\hat{c})}{(\hat{\tau}-\tau)^{3/2}} \right \rbrack.
\end{align}
In accordance with the Rohrlich-Teitelboim approach, it is proportional to the direct product of two null vectors $\hat{c}_{\mu}\hat{c}_{\nu}$ corresponding to the propagation of this part of field's energy-momentum exactly with the speed of light. Also, as was discussed above, the emitted part of the energy-momentum tensor depends on the entire history of the source's motion preceding the retarded proper time $\hat{\tau}$, as well as the radiation power determined by it.

\section{Electromagnetic radiation in five dimensions}

Having found the emitted part of the five-dimensional scalar field and the corresponding radiation energy-momentum tensor, let us now briefly discuss the electromagnetic and gravitational radiation in five dimensions. We start with the former.

\subsection{The setup}

Action of the electromagnetic field $A_{\mu}(x)$ interacting with the massive point particle moving along an arbitrary world line $z^{\mu}(\tau)$ in five dimensions is analogous to that of the scalar field \eqref{eq:sc_action} and has the form
\begin{equation}
\label{eq:em_action}
S = - \frac{1}{8\pi^2} \int d^{4+1}x \, F^{\mu\nu} F_{\mu\nu} - e \int d\tau \, \dot{z}^{\mu} A_{\mu} (z), \quad F_{\mu\nu} = \partial_{\mu} A_{\nu} - \partial_{\nu} A_{\mu},
\end{equation}
where $e$ is particle's electric charge. Here, we omit the particle's kinetic term in the action, given that its motion is completely governed by the external forces and, thus, the world line variables are non-dynamical.

The action \eqref{eq:em_action} yields the standard equation of motion of the electromagnetic field
\begin{align}
\label{eq:em_EoM}
&\square A^{\mu}(x) = 2\pi^2 j^{\mu}(x), \\
\label{eq:em_curr}
&j^{\mu}(x) = e \int d\tau \, \dot{z}^{\mu} \, \delta^{(4+1)} (x - z).
\end{align}
where we imposed on the field the Lorentz gauge condition to fix its gauge symmetry
\begin{equation}
\partial_{\mu} A^{\mu} = 0.
\end{equation}
Note that this condition requires the current to be conserved
\begin{equation}
\partial_{\mu} j^{\mu} = 0,
\end{equation}
and it is, when the observation point is off the world line, as can be easily seen from the Eq. \eqref{eq:em_curr}.

To determine the energy flux carried by the electromagnetic radiation we use symmetric energy-momentum tensor of the electromagnetic field
\begin{equation}
\label{eq:em_EMT}
\Theta_{\mu\nu} = \frac{1}{2\pi^2} \left \lbrack F_{\mu\,\cdot}^{\,\,\,\alpha} F_{\alpha\nu} + \frac{1}{4} \eta_{\mu\nu} F_{\alpha\beta} F^{\alpha\beta} \right \rbrack.
\end{equation}
By analogy with that of the scalar field, it is the bilinear functional of the field derivatives $\Theta \sim \partial A \partial A$ and, thus, admits defining the emitted part of the electromagnetic field derivative.

\subsection{Emitted part of the electromagnetic field}

By use of the Eqs. \eqref{eq:em_curr} and \eqref{eq:5D_Green} we obtain the retarded electromagnetic field of the point charge in form
\begin{equation}
A_{\mu}(x) = e \int d\tau \, \dot{z}_{\mu} \, \theta(X^{0}) \left \lbrack \frac{\delta(X^2)}{(X^2)^{1/2}} - \frac{1}{2} \frac{\theta(X^2)}{(X^2)^{3/2}} \right \rbrack.
\end{equation}

We calculate its derivative using the relation \eqref{eq:delta_int_parts}
\begin{multline}
\partial_{\nu} A_{\mu}(x) = e \int d\tau \, \theta(X^0) \left \lbrack \frac{3}{2} \frac{\theta(X^2)}{(X^2)^{5/2}} v_{\mu} X_{\nu} - \frac{\delta(X^2)}{(X^2)^{3/2}} v_{\mu} X_{\nu} - \right. \\ \left. - \frac{\delta(X^2)}{(vX)^2(X^2)^{1/2}} \lbrack (aX)-1 \rbrack v_{\mu} X_{\nu} + \frac{\delta(X^2)}{(vX)(X^2)^{1/2}} \lbrack a_{\mu} X_{\nu} - v_{\mu} v_{\nu} \rbrack \right \rbrack.
\end{multline}
Using the relations \eqref{eq:delta_compl_arg} and \eqref{eq:X_rho_expan}, and transforming the terms with delta functions by analogy with the Eqs. (\ref{eq:5D_sc_c-t_1}--\ref{eq:5D_sc_c-t_3}) we obtain the emitted part of the retarded electromagnetic field as the long-range part of its derivative with respect to the Lorentz-invariant distance
\begin{multline}
\label{eq:em_emit}
( \partial_{\nu} A_{\mu} )^{\rm rad} = \frac{e \hat{c}_{\nu}}{2^{5/2}\hat{\rho}^{3/2}} \int_{-\infty}^{\hat{\tau}} d\tau \left \lbrack \frac{3}{2} \frac{v_{\mu}}{(Z\hat{c})^{5/2}} - \frac{3}{2} \frac{\hat{v}_{\mu}}{(\hat{\tau} - \tau)^{5/2}} - \frac{(\hat{a}\hat{c})\hat{v}_{\mu} - \hat{a}_{\mu}}{(\hat{\tau} - \tau)^{3/2}} \right \rbrack.
\end{multline}
By analogy with the scalar field, in Eq. \eqref{eq:em_emit} all the physical information concerning the electromagnetic field in the wave zone is contained in the first term of the integrand, while the remaining ones just subtract the divergences contained in it at the upper integration limit.

Substituting the Eq. \eqref{eq:em_emit} into the symmetric energy-momentum tensor \eqref{eq:em_EMT} we arrive at the five-dimensional electromagnetic radiation energy-momentum tensor
\begin{align}
\label{eq:em_rad}
&\Theta_{\mu\nu}^{\rm rad} = - \frac{e^2 \hat{c}_{\mu} \hat{c}_{\nu}}{64 \pi^2 \hat{\rho}^3} {\cal A}_{\alpha}^{2}(x), \\
&{\cal A}_{\alpha}(x) = \int_{-\infty}^{\hat{\tau}} d\tau \left \lbrack \frac{3}{2} \frac{v_{\alpha}}{(Z\hat{c})^{5/2}} - \frac{3}{2} \frac{\hat{v}_{\alpha}}{(\hat{\tau} - \tau)^{5/2}} - \frac{(\hat{a}\hat{c})\hat{v}_{\alpha} - \hat{a}_{\alpha}}{(\hat{\tau} - \tau)^{3/2}} \right \rbrack.
\end{align}
The obtained tensor structure of the radiated part of the energy-momentum tensor corresponds to the propagation of the associated energy flux exactly with the speed of light. Also, by analogy with the scalar field \eqref{eq:sc_EMT_emit_1}, it depends on the history of charge's motion preceding the retarded proper time.

\section{Five-dimensional gravitational radiation}

Now we turn to the calculation of the gravitational radiation produced by the point particles.  If interaction between them is also gravitational, the problem is not described by the linearized approximation, and the quadratic terms in expansion of the Einstein tensor are required to take into account the contribution of field stresses to radiation \cite{Weinberg1972}. If the dominant forces are non-gravitational, the corresponding field stresses are also required, making the calculation rather difficult. Here we calculate only the local contribution of particles themselves, so the result is incomplete. We give it just to reveal difference with the scalar and electromagnetic cases and to show how to deal with polarisations.

\subsection{The setup}

We start with the generally covariant Einstein-Hilbert action for the five-dimensional gravity interacting with massive point particle and some external field governing its motion
\begin{equation}
S = \frac{1}{2\kappa_5} \int d^{4+1}x \sqrt{-g} R - m \int d\tau \sqrt{g_{\alpha\beta}(z) \dot{z}^{\alpha} \dot{z}^{\beta}} + S_{\rm F},
\end{equation}
where $\kappa_5$ is the five-dimensional gravitational constant, $g$ is the determinant of metric tensor, and $S_{\rm F}$ is the action for the external field moving point particle.

We linearise the Einstein's equation over the background Minkowski metric
\begin{equation}
g_{\mu\nu}(x) = \eta_{\mu\nu} + h_{\mu\nu}(x), \quad |h_{\mu\nu}| \ll 1.
\end{equation}
Introducing the reduced metric perturbations
\begin{equation}
\bar{h}_{\mu\nu} = h_{\mu\nu} - \frac{1}{2} \eta_{\mu\nu} h, \quad h = \eta^{\alpha\beta}h_{\alpha\beta},
\end{equation}
we arrive at the linearised Einstein equation
\begin{align}
\label{eq:lin_Ein_eq}
&\square \bar{h}_{\mu\nu}(x) = 2 \kappa_{5} \left( T_{\mu\nu}^{\rm P}(x) + T_{\mu\nu}^{\rm F}(x) \right), \\
\label{eq:pp_EMT}
&T_{\mu\nu}^{\rm P}(x) = m \int d\tau \, \dot{z}_{\mu} \dot{z}_{\nu} \, \delta^{(5)} (x - z), \quad T_{\mu\nu}^{\rm F}(x) = - \left. \frac{2}{\sqrt{-g}} \frac{\delta S_{\rm F}}{\delta g^{\mu\nu}} \right \vert_{g=\eta}
\end{align}
where $T_{\mu\nu}^{\rm P}(x)$ is the energy-momentum of point particle moving on the flat Minkowski background and interacting with some external field, which energy-momentum tensor is given by $T_{\mu\nu}^{\rm F}(x)$. To obtain the Eq. \eqref{eq:lin_Ein_eq} we, also, imposed the Lorentz gauge condition on the metric perturbations
\begin{equation}
\partial^{\mu} \bar{h}_{\mu\nu} = 0,
\end{equation}
to fix the gauge symmetry of the system. Such a condition requires the particle's and external field's energy momentum-tensors to be jointly conserved
\begin{equation}
\partial^{\mu} \left( T_{\mu\nu}^{\rm P} + T_{\mu\nu}^{\rm F} \right) = 0,
\end{equation}
which is assumed to be valid on-shell.

To determine the energy-momentum flux carried by the gravitational radiation we use the effective energy-momentum tensor of metric perturbations, by analogy with \cite{maggiore2008},
\begin{align}
\label{eq:gw_EMT}
&t_{\mu\nu} = \frac{1}{4\kappa_{5}} \left \langle \partial_{\mu} \bar{h}_{ij}^{\rm TT} \, \partial_{\nu} \bar{h}_{ij}^{\rm TT} \right \rangle, \quad \bar{h}_{ij}^{\rm TT}(x) = \Lambda_{ij,kl}(\mathbf{n}) \bar{h}_{kl}(x), \\
\label{eq:L_ten}
&\Lambda_{ij,kl} (\mathbf{n}) = P_{ik}P_{jl} - \frac{1}{3} P_{ij} P_{kl}, \quad P_{ij} = \delta_{ij} - n_{i}n_{j}, \quad n^{i} = \frac{x^i}{r},
\end{align}
where we assume the particle's motion to be periodic, and $\langle \ldots \rangle$ is the averaging over the period. Here, we turned into the transverse-traceless gauge by the contraction of metric perturbations \eqref{eq:gw_EMT} with the projector \eqref{eq:L_ten} defined by analogy with \cite{maggiore2008}. Note that differentiation of the projector would increase the fall-off asymptotic of the field, so its derivative can be neglected when one computes the emitted part of the gravitational field in the transverse-traceless gauge.

\subsection{Emitted part of the gravitational field (incomplete)}

Using the Eqs. \eqref{eq:5D_Green} and \eqref{eq:pp_EMT} we find the retarded gravitational field of the point particle as
\begin{equation}
\bar{h}_{\mu\nu}^{\rm P}(x) = \frac{m \kappa_{5}}{\pi^2} \int d\tau \, \dot{z}_{\mu} \dot{z}_{\nu} \, \theta(X^{0}) \left \lbrack \frac{\delta(X^2)}{(X^2)^{1/2}} - \frac{1}{2} \frac{\theta(X^2)}{(X^2)^{3/2}} \right \rbrack.
\end{equation}

We compute its derivative integrating by parts by use of the relation \eqref{eq:delta_int_parts} arriving at
\begin{multline}
\label{eq:gw_deriv}
\partial_{\alpha} \bar{h}_{\mu\nu}^{\rm P} = \frac{m \kappa_{5}}{\pi^{2}} \int d\tau \, \theta(X^{0}) \left \lbrack \frac{3}{2} \frac{\theta(X^{2})}{(X^2)^{5/2}} v_{\mu} v_{\nu} X_{\alpha} - \frac{\delta(X^{2})}{(X^2)^{3/2}} v_{\mu} v_{\nu} X_{\alpha} - \right. \\ - \left. \frac{\delta(X^{2})}{(vX)^{2} (X^2)^{1/2}} \lbrack (aX) - 1 \rbrack v_{\mu} v_{\nu} X_{\alpha} + \frac{\delta(X^{2})}{(vX) (X^2)^{1/2}} \lbrack 2 a_{(\mu} v_{\nu)} X_{\alpha} - v_{\mu} v_{\nu} v_{\alpha} \rbrack \right \rbrack,
\end{multline}
where $a^{\mu}=d^{2}z^{\mu}/d\tau^{2}$ is the acceleration vector, and we define the symmetrisation over two indices as $A_{(\mu} B_{\nu)}=(A_{\mu} B_{\nu} + A_{\nu} B_{\mu})/2$.

Extracting the long-range part of the gravitational field derivative by use of the relations \eqref{eq:delta_compl_arg} and \eqref{eq:X_rho_expan} and transforming the terms with delta functions by analogy with Eqs. (\ref{eq:5D_sc_c-t_1}--\ref{eq:5D_sc_c-t_3}) we obtain the emitted part of the gravitational field as
\begin{equation}
\label{eq:gw_emit}
\left( \partial_{\alpha} \bar{h}_{\mu\nu}^{\rm P} \right)^{\rm rad} = \frac{m \kappa_{5} \hat{c}_{\alpha}}{2^{5/2} \pi^{2} \hat{\rho}^{3/2}} \int_{-\infty}^{\hat{\tau}} d\tau \left \lbrack \frac{3}{2} \frac{v_{\mu} v_{\nu}}{(Z\hat{c})^{5/2}} - \frac{3}{2} \frac{\hat{v}_{\mu} \hat{v}_{\nu}}{(\hat{\tau} - \tau)^{5/2}} - \frac{(\hat{a}\hat{c}) \hat{v}_{\mu} \hat{v}_{\nu} - 2 \hat{a}_{(\mu} \hat{v}_{\nu)}}{(\hat{\tau} - \tau)^{3/2}} \right \rbrack.
\end{equation}
By analogy with the scalar \eqref{eq:5D_sc_emit} and electromagnetic \eqref{eq:em_emit} fields, it consists of one integral determining the properties of the gravitational field in the wave zone and two integral being counter-terms eliminating the divergences from the first one.

Substituting the obtained expression for the emitted part of the retarded gravitational field \eqref{eq:gw_emit} into the Eq. \eqref{eq:gw_EMT} we find the energy-momentum tensor of gravitational radiation generated by the point particle
\begin{align}
&\left( t_{\mu\nu}^{\rm P} \right)^{\rm rad} = \frac{m^2 \kappa_{5} \hat{c}_{\mu} \hat{c}_{\nu}}{128 \pi^4 \hat{\rho}^{3}} {\cal A}_{ij}^{\rm TT} (x) {\cal A}_{ij}^{\rm TT} (x), \quad {\cal A}_{ij}^{\rm TT} (x) \equiv \Lambda_{ij,kl} (\mathbf{n}) {\cal A}_{kl} (x), \\
& {\cal A}_{ij}(x) = \int_{-\infty}^{\hat{\tau}} d\tau \left \lbrack \frac{3}{2} \frac{v_{i} v_{j}}{(Z\hat{c})^{5/2}} - \frac{3}{2} \frac{\hat{v}_{i} \hat{v}_{j}}{(\hat{\tau} - \tau)^{5/2}} - \frac{(\hat{a}\hat{c}) \hat{v}_{i} \hat{v}_{j} - 2 \hat{a}_{(i} \hat{v}_{j)}}{(\hat{\tau} - \tau)^{3/2}} \right \rbrack.
\end{align}
As in the cases of scalar \eqref{eq:sc_EMT_emit_1} and electromagnetic \eqref{eq:em_rad} radiation, the emitted part of the gravitational field energy-momentum tensor has the tensor structure corresponding to the propagation of the associated energy-momentum flux exactly at the speed of light and depends on the entire history of the particle's motion preceding the retarded proper time $\hat{\tau}$.

\section{Conclusions}

In this short article, we have demonstrated that despite the Huygens principle violation in odd spacetime dimensions, radiation can be computed by the integration of the energy-momentum flux in the wave zone. However, it requires the Lorentz-invariant modification of the radiation and wave zone definitions, in accordance with the Rohrlich-Teitelboim approach. Also, due to the Huygens principle violation, the emitted part of the field depends on the history of the source's motion preceding the retarded proper time, in contrast with even dimensions, where properties of radiation at given moment of time are determined only by the state of the source at the retarded time. Another feature of the odd dimensions is that the retarded field is given by the sum of separately divergent integrals. Nevertheless, these divergence mutually cancel out and the resulting field is finite.

Based on the Rohrlich-Teitelboim approach, we have considered the radiation of the scalar, electromagnetic and gravitational fields by point particle in five dimensions. We computed the emitted parts of the fields' energy-momentum tensors. The obtained expressions have the tensor structures corresponding to the propagation of the radiated energy-momentum in space exactly with the speed of light. Also, they are given by the sum of separately divergent integrals over the history of particle's motion, one of which carries all the physical information concerning the field in the wave zone, while the remaining ones are just the counter-terms subtracting the divergences from the former. To make the convergence of the resulting tensor explicit, one has to integrate the first term by parts.

In our previous work \cite{Galtsov:2020hhn}, we have considered the scalar synchrotron radiation in three and five dimensions. The results were checked by the calculation of the spectral decompositions of the radiation power, which are indifferent to the dimensionality of the spacetime. Based on this, we assume that the obtained similar expressions for the emitted parts of the electromagnetic and gravitational fields should, also, be valid.

\section*{Acknowledgements}

The work of M. Kh. was supported by the “BASIS” Foundation Grant No. 20-2-10-8-1. The work of D.G. was supported by the Russian Foundation for Basic Research on the project 20-52-18012, and the Scientific and Educational School of Moscow State University "Fundamental and Applied Space Research".

\end{document}